\documentclass{article}
\usepackage[english]{babel}
\usepackage[a4paper,top=2cm,bottom=2cm,left=3cm,right=3cm,marginparwidth=1.75cm]{geometry}

\usepackage[utf8]{inputenc}
\usepackage{pgfplots}
\DeclareUnicodeCharacter{2212}{−}
\usepgfplotslibrary{groupplots,dateplot}
\usetikzlibrary{patterns,shapes.arrows}
\pgfplotsset{compat=newest}
\usepackage{amsmath}
\usepackage{cite}
\usepackage{graphicx}
\usepackage{amssymb}
\usepackage{subfigure}
\usepackage[colorlinks=true, allcolors=black]{hyperref}
\usepackage{import}
\usepackage{authblk}

\title{Ultra-low Frequency Noise External Cavity Diode Laser Systems for Quantum Applications}

\author[1,2,*]{Niklas Kolodzie}
\author[1]{Ivan Mirgorodskiy}
\author[1]{Christian Nölleke}
\author[2,3]{Piet O. Schmidt}

\affil[1]{TOPTICA Photonics AG, Lochhamer Schlag 19, D-82166 Gräfelfing, Germany}
\affil[2]{Physikalisch-Technische Bundesanstalt, Bundesallee 100, D-38116 Braunschweig, Germany}
\affil[3]{Institut für Quantenoptik, Leibniz Universität Hannover, D-30167 Hannover, Germany}
\affil[*]{Corresponding author: niklas.kolodzie@toptica.com}

\begin{document}
\maketitle


\begin{abstract}
We present two distinct ultra-low frequency noise lasers at $729$\,nm with a fast frequency noise of $30$\,Hz$^{2}$/Hz, corresponding to a Lorentzian linewidth of $0.1$\,kHz. The characteristics of both lasers, which are based on different types of laser diodes, are investigated using experimental and theoretical analysis with a focus on identifying the advantages and disadvantages of each type of system. Specifically, we study the differences and similarities in mode behaviour while tuning frequency noise and linewidth reduction. Furthermore, we demonstrate the locking capability of these systems on medium-finesse cavities. The results provide insights into the unique operational characteristics of these ultra-low noise lasers and their potential applications in quantum technology that require high levels of control fidelity.
\end{abstract}


\section{Introduction} \label{1}

Continuous-wave single longitudinal mode lasers are a critical component in a variety of quantum applications, such as optical clocks \cite{Wieman1991, Ludlow2015, Pavone1996, Udem2002, Stoehr2006, Fasano2021}, quantum computing \cite{Wang2001, Leibfried2003, Haeffner2008, Blatt2008, Monz2016, Pogorelov2021} and quantum communication \cite{Kimble2008, Ritter2012, Hanson2018}. These applications require increasingly high levels of precision and robustness, which impose stringent requirements on the frequency noise (FN) of the laser. In particular, it is crucial to keep FN at a minimum level over a wide range of Fourier frequencies: slow FN (DC - $5$\,kHz) contributes to long-term stability, while fast FN ($>5$\,kHz) ultimately limits qubit coherence times and gate fidelities \cite{Jiang2023, Nakav2023}.

Optically pumped solid state lasers based on titanium sapphire (TiSa) are a frequent choice for the mentioned applications. TiSa lasers offer several advantages over other types of lasers, such as high power outputs and low levels of fast FN \cite{Muller2006, Freund2023}. These systems can be tuned over a wide range of wavelengths, covering the visible and near-infrared spectrum. In addition to their positive aspects, it should be noted that TiSa systems also have certain drawbacks. These include the demand for a significant amount of physical space, high energy consumption and a need for frequent maintenance. As a result, these systems may not be suitable for highly integrated and flexible setups, which are enablers for the modern quantum market \cite{Stuhler2021, Pogorelov2021, Moses2023}. For particular experiments fiber-based laser systems are also a possible solution: they provide low FN behaviour and high output powers, but they are only available for specific wavelengths \cite{Dietrich2009, Brewer2019, Kraus2022, Sarkar2022}.

\begin{figure}[t!]
\centering
    \subfigure{\includegraphics[height=0.33\textwidth]{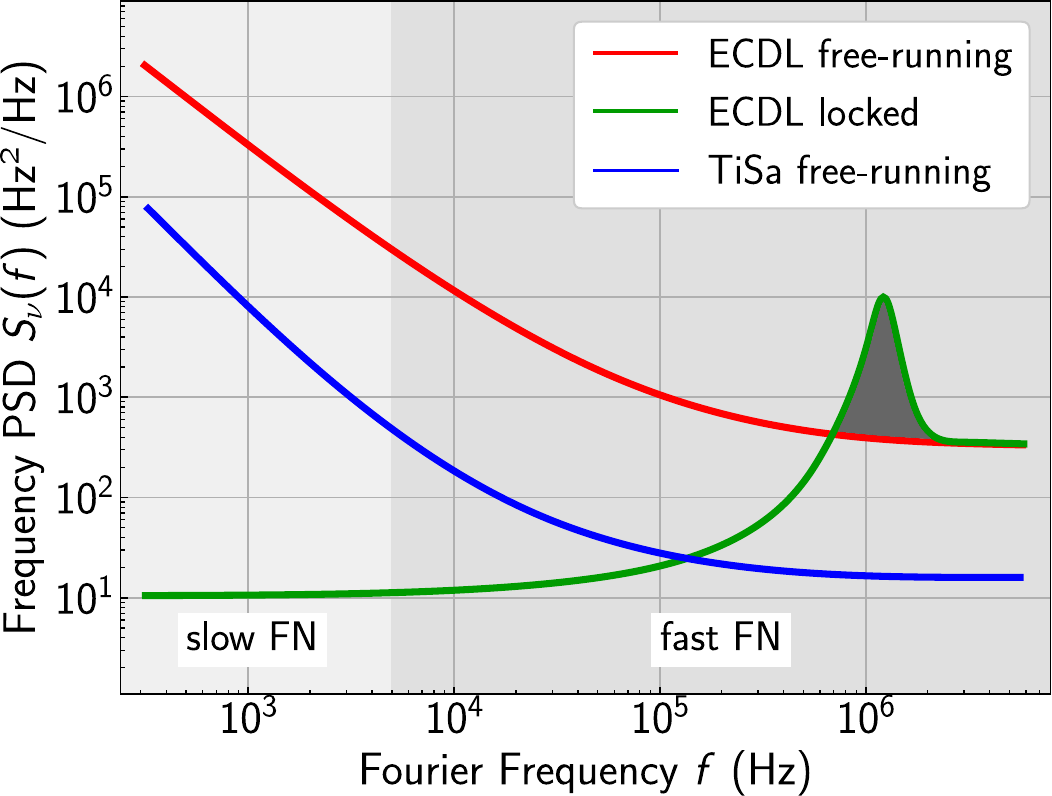}}
    \caption{\label{fig:servo}Typical FN behaviour of a locked and a free-running ECDL \cite{Stoehr2006} and a free-running TiSa \cite{Senko2022}. Compared to the free running TiSa, the locked ECDL shows a smaller FN for slower Fourier frequencies. The excess noise (dark grey area) is shifted to faster FN, leading to a servo bump. The limit of FN reduction for fast Fourier frequencies is determined by the laser system and the bandwidth of the control loop.}
\end{figure}

The external cavity diode laser (ECDL) is a versatile laser concept based on a semiconductor gain medium in the form of a laser diode \cite{Ricci1995, Henry1987, Goldberg1982, Saliba2009}. They are widely used in various applications due to their small size, low cost and the ability to cover many wavelengths. Compared to TiSa systems, ECDLs have higher levels of FN due to the shorter laser cavity, which can limit the fidelity of coherent manipulations in experiments \cite{Senko2022}. To enable the usage of ECDLs for high-precision applications, it is necessary to actively or passively minimize the FN in a wide range of Fourier frequencies. One common technique for reducing the FN is frequency stabilization using an active electronic feedback signal, generated by comparing the laser frequency with a frequency reference \cite{Nazarova2006, Stoehr2006, Ludlow2007}. This so-called frequency lock of a laser can track and correct rapid fluctuations in the laser frequency. The frequency control loop pushes the frequency noise to the Fourier frequencies outside its bandwidth, as schematically shown in Figure \ref{fig:servo}. This leads to high excess noise in the fast FN domain - this effect is known as a servo bump. The stabilization technique, the choice of the controller parameters and the overall bandwidth of the control loop determine the location and the amplitude of the servo bump. The maximum achievable value of the bandwidth is limited by the physics of frequency tuning in the semiconductor material of the laser diode to a few MHz \cite{Coldren2012}. Typically, frequency stabilization inevitably results in an excess of fast FN around the bandwidth frequency. Far above the lock bandwidth, the FN is not affected and stays at the level of the free-running laser. For each application, a compromise must be made between reducing slow FN and creating excessive fast FN.

Various concepts have been proposed to reduce FN in ECDLs passively without generation of fast excess FN. One such approach is to filter light of a diode-based laser with an optical cavity and to seed another laser diode \cite{Hald2005, Labaziewicz2007, Nazarova2008, Akerman2015}. The optical cavity works as a low-pass filter for phase noise above the cavity linewidth. The spectral properties of filtered light are transferred to the seeded laser diode. However, the transmitted power is often in the \textmu W regime and too low for a direct seeding of, e.g., a tapered amplifier. Consequently, multiple amplification stages are required to achieve several milliwatts of output power. This makes the laser design complex, large and expensive. Another concept to reduce FN in ECDLs is resonant optical feedback from an optical cavity back into the laser diode \cite{Dahmani1987, Doringshoff2007, Zhao2011, Zhao2012, Lewoczko-Adamczyk2015, King2018, Krinner2023}. In this setup the feedback cavity is an extension of the laser cavity. The extent of FN reduction depends on the finesse ($\mathcal{F}$) of the cavity. To satisfy the strict resonance condition, active stabilization methods are necessary. For a higher passive stability, the whole setup must be decoupled from the environment and has to be as small as possible.

A simplified version of this idea is to extend the ECDL cavity with an additional long optical delay line instead of using a separate optical cavity. The larger photon lifetime in the laser can lead to a FN reduction by several orders of magnitude. One convenient technical implementation is to use an optical fiber as an optical delay line \cite{Lin2012, Samutpraphoot2014, Yamoah2019}. However the drawback of this approach is that due to the extension of the laser cavity, the free spectral range (FSR) is reduced to several MHz. This leads to a smaller longitudinal mode spacing, which makes the system more vulnerable to mode hops. This approach was primarily implemented for distributed feedback laser diodes, which are more robust against environmental changes such as temperature and pressure as well as possess a high mode-hope-free tuning range. However, these systems are only available for a small number of wavelengths in the near-infrared spectrum. To build Fabry-Pérot (FP) and anti-reflection coated (AR) laser diode-based ultra-low noise lasers (ULNL) for a wide range of wavelengths, the robustness of systems with laser diodes must be significantly increased.

In this study, we analyze the behaviour of ULNL based on an ECDL with weak optical feedback using an additional fiber cavity at $729$\,nm. This wavelength is particularly relevant for coherent manipulation of calcium ions that are utilized as optical clocks or qubits in quantum information processing and quantum simulations. These applications have strict FN requirements in the MHz range. Specifically, we investigate the influence of the light source on the system behaviour by comparing two identical laser setups, differing only in the type of the used laser diode. Our results demonstrate that the use of an FP laser diode as a source significantly improves the system's behaviour in the form of mode stability and frequency selection. However, when utilizing a refined FP laser diode with an AR coating on the front facet \cite{Serenyi2001}, the system becomes much more stable in terms of single-mode behaviour. These findings suggest that careful consideration of the type of laser diode used is critical for achieving stable and reliable operation of ULNL systems with weak optical feedback. This work provides valuable insights for the development and optimization of such systems for a range of applications in the field of ECDLs. In Section \ref{2} we introduce the theory for our work, in Section \ref{3} we describe our experimental setup including the measurement and characterisation system and in Section \ref{4} we present the theoretical and experimental results for our ULNLs.


\section{Theory} \label{2}

\begin{figure}[h!]
\centering
    \subfigure{\includegraphics[height=0.22\textwidth]{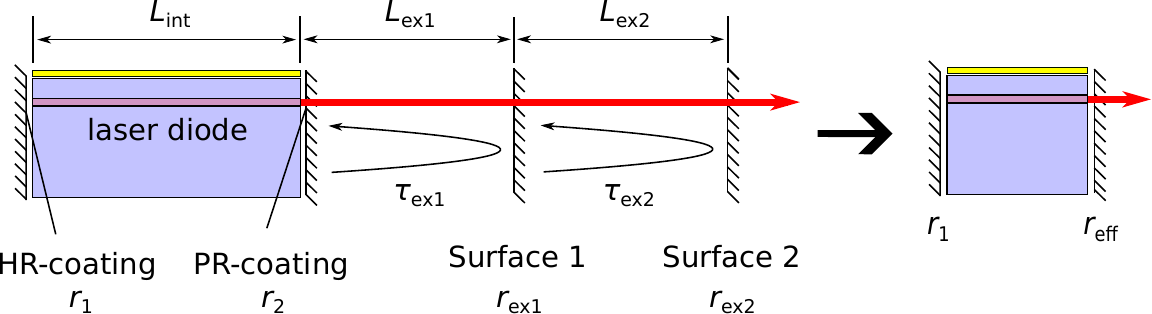}}
    \caption{\label{fig:scheme}Schematic of the multi-mirror model for laser including laser diode and the two external cavities. The system can be reduced to a two-mirror laser with $r_{1}$ and $r_{\mathrm{eff}}$.}
\end{figure}

For our theoretical consideration we look at two different topics: On the one hand we describe the ULNL in terms of FN and linewidth reduction, on the other hand we focus on the mode structure of the two lasers with different types of LDs. Figure \ref{fig:scheme} presents schematically a laser based on a laser diode with two additional external cavities. The whole laser consists of four reflective surfaces resulting in three coupled cavity parts: The active medium of the laser diode (amplitude reflection coefficients $r_{1}$ and $r_{2}$ with the refractive-index-dependent cavity length $n_{\mathrm{LD}}\cdot L_{\mathrm{int}}$) forms the first cavity. The second part ($r_{\mathrm{ex1}}$ with $L_{\mathrm{ex1}}$) creates the ECDL with the frequency-selective grating \cite{Saliba2009}. The long additional fiber with an end-mirror ($r_{\mathrm{ex2}}$ with $L_{\mathrm{ex2}}$) completes the ULNL. All relevant parameters are listed in Table \ref{tab:widgets}.

\subsection*{Frequency noise and linewidth reduction}

For the analysis of the FN of lasers we determine the power spectral density $S_{\nu}(f)$ over a wide range of Fourier frequencies from $300$\,Hz to $6$\,MHz, measured by using a delayed-self-heterodyne method, explained in Section \ref{3} \cite{Smith2007, Tsuchida2011, Michaud-Belleau2016, Schmidt-Eberle2023}. Since we are interested in the fast FN regime, we extract the Lorentzian linewidth from the FN spectrum. The Lorentzian linewidth of a laser stems from the inherent white noise generated by spontaneous emission. It is a spectral purity parameter, also known as the intrinsic linewidth. The technical noise components of the laser, which can arise due to temperature fluctuations or mechanical or acoustic vibrations, are not taken into account in this definition. We employ the approach discussed in \cite{Halford1972, Hjelme1991} to convert fast FN to the Lorentzian linewidth $\Delta\nu_{L}$. Therefore, we average over the frequency range of $200$\,kHz to $6$\,MHz, where $S_{\nu}(f)$ of our lasers becomes nearly flat. In this range the white noise level $S_{\nu 0}$ originating from the spontaneous emission is reached. The fast Lorentzian linewidth $\Delta\nu_{L}$ can be calculated by
\begin{equation}
\label{eq:5}
\Delta\nu_{L} = \pi S_{\nu 0}.
\end{equation}
Additional optical feedback back to the ECDL decreases the Lorentzian linewidth $\Delta\nu_{L,\mathrm {ECDL}}$ according to Schawlow-Townes theory \cite{Townes1958}. The influence of this feedback is described in \cite{Li1989, Li-Telle1989, Laurent1989}: An additional cavity part with an extra length leads to an extended photon lifetime in the system. To calculate the reduced linewidth $\Delta\nu_{L, \mathrm{ULNL}}$, we use the equation
\begin{equation}
\label{eq:6}
\Delta\nu_{L, \mathrm{ULNL}} = \frac{\Delta\nu_{L,\mathrm {ECDL}}}{\bigg(1+\sqrt{1+\alpha_{H}^2}\sqrt{\beta} \dfrac{L_{\mathrm{ex2}}}{n_{\mathrm{LD}}\cdot L_{\mathrm{int}}+L_{\mathrm{ex1}}}\bigg)^2}
\end{equation}
with the linewidth enhancement factor $\alpha_{H}$ (relates phase changes to changes of the gain) \cite{Henry1982}, the separate cavity lengths $n_{\mathrm{LD}}\cdot L_{\mathrm{int}}$, $L_{\mathrm{ex1}}$ and $L_{\mathrm{ex2}}$ and the feedback ratio $\beta$, defined as the ratio between feedback power and output power. The latter is responsible for both the linewidth reduction and the stability of the system. To investigate the influence of $\beta$ on $\Delta\nu_{L, \mathrm{ULNL}}$, we can change this value in our laser experimentally over a wide range, described in Section \ref{4}. Different operation regimes for variable feedback ratios $\beta$ are described in \cite{Tkach1986, Okoshi1980, Kane2005, Coldren2012}: A too-strong feedback ratio $\beta$ leads to multiple allowed lasing mode solutions and turns single-mode behaviour to multi-mode. This effect is known as coherence collapse, which will be analysed in Section \ref{4}.

\begin{table}[t!]
\centering
\begin{tabular}{l|c|c|c}
\textbf{Parameter} & \textbf{Symbol} & \textbf{Value} & \textbf{Unit} \\ [1mm]
\hline
Frequency & $\nu$ & $411.2$ & THz \\
Linewidth enhancement factor & $\alpha_{H}$ & $4\leq\alpha_{H}\leq6$ &  \\
Laser diode length & $L_{\mathrm{int}}$ & $1.2 \cdot 10\textsuperscript{-3}$ & m \\
Laser diode refractive index & $n_{\mathrm{LD}}$ & $3.5$ & \\
Laser diode reflection coefficient rear facet & $r_{\mathrm{1}}^2$ & $0.99$ & \\
Laser diode reflection coefficient front facet & $r_{\mathrm{2, FP}}^2$ & $0.05$ & \\
 & $r_{\mathrm{2, AR}}^2$ & $0.001$ & \\
First external cavity length & $L_{\mathrm{ex1}}$ & $38 \cdot 10\textsuperscript{-3}$ & m \\
First external cavity reflection coefficient & $r_{\mathrm{ex1}}^2$ & $0.15$ & \\
Second external cavity length & $L_{\mathrm{ex2}}$ & $2.55$ & m \\
Second external cavity feedback ratio & $\beta$ & $-47 < \beta < -27$ & dB \\
\end{tabular}
\caption{\label{tab:widgets}Laser parameters used in the simulations.}
\end{table}

\subsection*{Multi-mirror laser cavity and mode stability}
For analyzing the single longitudinal mode behaviour of an ULNL, it is necessary to focus on each cavity mode of each part of the laser system. The simplest cavity of a laser consists of two reflective surfaces in a defined distance. This concept is known as the two-mirror model. If more cavity interfaces such as partially reflective surfaces are involved, the scattering matrix formalism can be used to describe the multi-mirror laser cavity \cite{Henry1987}. This technique allows to reduce a configuration with several coupled cavities back to a two-mirror system with modified properties, without loss of generality.

To reduce the multi-mirror system to a two-mirror system, we introduce the definition of the effective amplitude reflection $r_{\mathrm{eff}}$ \cite{Coldren2012}. This complex number combines the diode front-facet and two external cavity surfaces in one value using scattering theory assuming no losses and no dispersion:
\begin{equation}
    \label{eq:1}
    r_{\mathrm{eff}} = S_{11} + \frac{S_{12} S_{21} r_{\mathrm{ex}2}\, e^{-i\omega \tau_{\mathrm{ex}2}}}{1 - S_{22} r_{\mathrm{ex}2}\, e^{-i\omega \tau_{\mathrm{ex}2}}}
\end{equation}
with the amplitude reflection coefficient of the second external cavity $r_{\mathrm{ex}2}$, the photon lifetime in the second external cavity $\tau_{\mathrm{ex}2} = 2 L_{\mathrm{ex}2} / c_{0}$ and the oscillation laser frequency $\omega = 2 \pi \nu$. The corresponding scattering coefficients $S_{ij}$ are
\begin{equation*}
    S_{11} = r_{2} + \frac{t_{2}^2 r_{\mathrm{ex1}}\, e^{-i\omega \tau_{\mathrm{ex1}}}}{1 + r_{2} r_{\mathrm{ex1}}\, e^{-i\omega \tau_{\mathrm{ex1}}}},\,\,\, S_{12} = S_{21} = \frac{t_{2} t_{\mathrm{ex1}}\, e^{-\frac{1}{2} i\omega \tau_{\mathrm{ex1}}}}{1 + r_{2} r_{\mathrm{ex1}}\, e^{-i\omega \tau_{\mathrm{ex1}}}},\,\,\, S_{22} = -r_{\mathrm{ex1}} - \frac{t_{\mathrm{ex1}}^2 r_{2}\, e^{-i\omega \tau_{\mathrm{ex1}}}}{1 + r_{2} r_{\mathrm{ex1}}\, e^{-i\omega \tau_{\mathrm{ex1}}}}
\end{equation*}
with the amplitude transmission coefficients $t_{2}$ and $t_{\mathrm{ex1}}$ and $t_{i}^2 + r_{i}^2 = 1$, the amplitude reflection coefficient of the ECDL $r_{\mathrm{ex1}}$ and the photon lifetime in the ECDL cavity $\tau_{\mathrm{ex1}} = 2 L_{\mathrm{ex1}} / c_{0}$. In summary, the ULNL is reduced to a laser cavity consisting of two mirrors with reflectivities $r_{1}$ and $r_{\mathrm{eff}}$.

The distinction between an FP laser diode and an AR laser diode is the partial reflection of the laser diode front facet. For an FP laser diode the typical reflectivity of a front facet is $r_{2, \mathrm{FP}}^2\approx0.05$. To reduce this value and thus suppress the internal modes, an AR coating can be applied. This coating reduces the partial reflection down to $r_{2, \mathrm{AR}}^2\approx0.001$ \cite{Serenyi2001}. Compared to diode-based systems with an FP laser diode, a reduced amplitude reflection coefficient of the front facet of the laser diode has less influence on the mode behaviour of the whole system. An AR coating strongly suppresses the inner longitudinal modes of the laser diode, while an FP laser diode adds one additional influencing surface to the cavity structure, which leads to more a complex mode structure.

The resulting emission frequency $\nu$ of an ULNL depends on the longitudinal mode behaviour of all included cavities. After converting the complex multi-mirror system to an effective two-mirror system it is possible to calculate the mode structure using the classical formula for the FP cavity transmission \cite{Saleh1991}:
\begin{equation}
\label{eq:2}
T(\nu) = \frac{T_{\mathrm{max}}}{1+\bigg(\dfrac{2\mathcal{F}}{\pi}\bigg)^2\mathrm{sin}^2\bigg(\dfrac{\pi\nu}{{\mathrm{FSR}}}\bigg)}
\end{equation}
with the maximum transmittance 
\begin{equation}
\label{eq:3}
T_{\mathrm{max}} = \frac{(1-\lvert r_{i} \lvert^2)(1-\lvert r_{j} \lvert^2)}{(1-\lvert r_{i} r_{j} \lvert)^2}
\end{equation}
and the finesse
\begin{equation}
\label{eq:4}
\mathcal{F} = \frac{\pi \sqrt{\lvert r_{i} r_{j}  \lvert}}{1-\lvert r_{i}r_{j}  \lvert}
\end{equation}
where $r_{i}$ and $r_{j}$ are the reflectivities of the involved cavity mirrors and the $\mathrm{FSR} = c / 2 n_{i} L_{i}$ are used. 

\begin{figure}[b!]
\centering
    \subfigure[\label{fig:modesa}]{\includegraphics[height=0.33\textwidth]{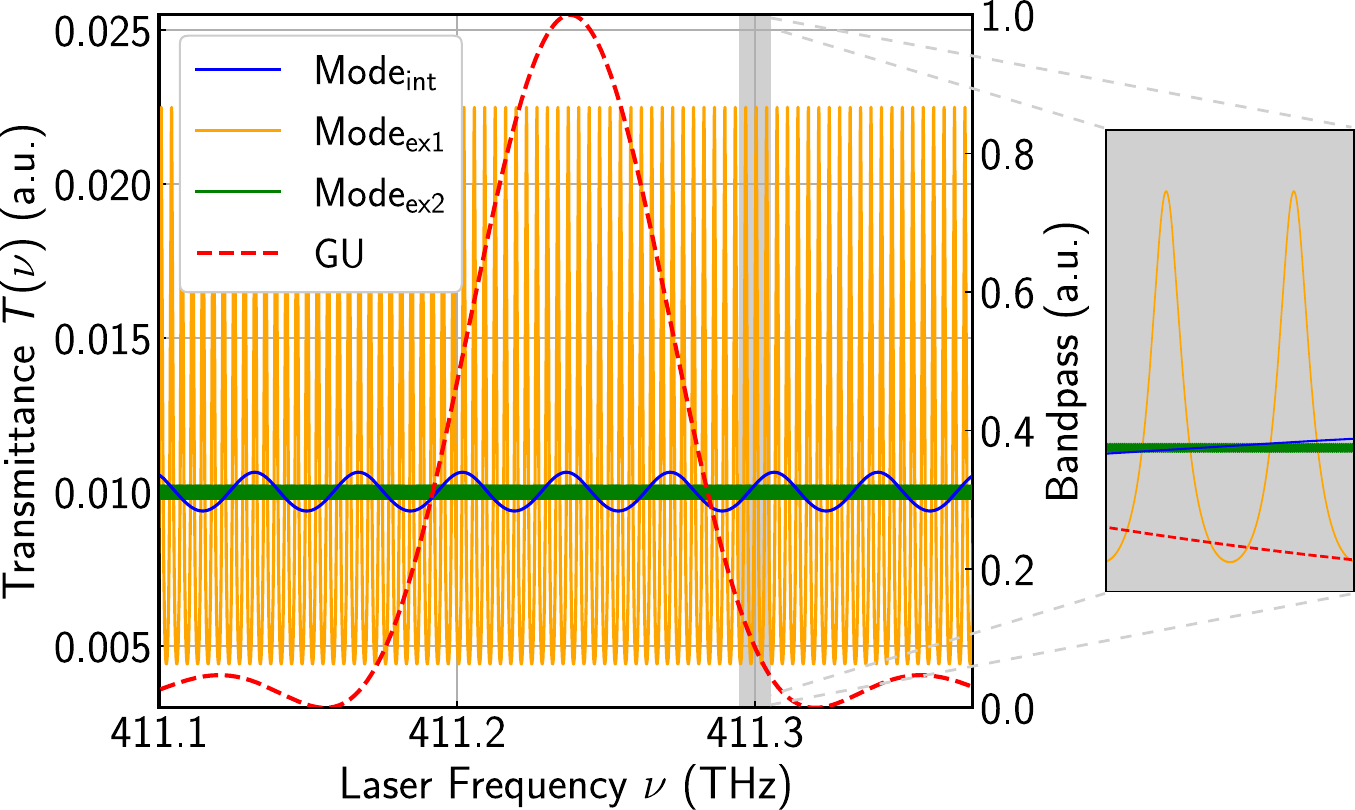}}\qquad
    \subfigure[\label{fig:modesb}]{\includegraphics[height=0.33\textwidth]{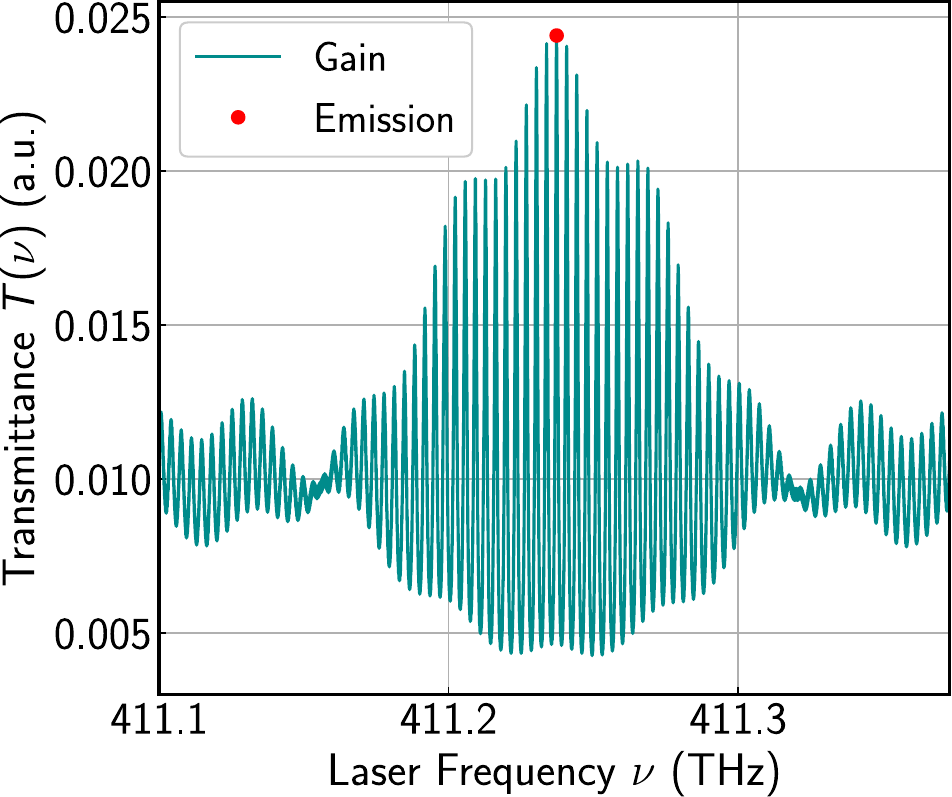}}\\
    \caption{\label{fig:modes}Simulated $T(\nu)$ for an ULNL with an AR laser diode. (a) The different longitudinal mode structures for all involved cavities: The internal (blue) and the two external (orange and green) cavity modes are calculated while using the real laser parameters for the AR laser diode case. The dashed line describes the frequency selective bandpass function of the grating (GU). Note: $\mathrm{Mode}_{\mathrm{ex2}}$ has such a high frequency that the structure cannot be resolved in the figure. (b) Combined mode structure of an ULNL calculated with $r_{1}$ and $r_{\mathrm{eff}}$.}
\end{figure}

To demonstrate the whole concept of simulating a mode structure of an ULNL, we start in Figure \ref{fig:modesa} with the transmission spectra for each cavity part separately. A system with an AR laser diode is shown. The mode structure of the internal laser diode $\mathrm{Mode}_{\mathrm{int}}$ is shown in blue. The relatively short cavity has an FSR $\approx35$\,GHz. The two external cavity parts are longer, so the FSR of both cavities is smaller ($\mathrm{Mode}_{\mathrm{ex1}}$ in yellow and $\mathrm{Mode}_{\mathrm{ex2}}$ in green). The smallest FSR is given by the length of the fiber cavity (FSR $\approx55$\,MHz). The amplitude of the cavity transmittance $T(\nu)$ depends on the reflectivity of the surface. Additionally we show the bandpass profile of the frequency-selective grating, which is defined by the groove profile and the number of illuminated lines \cite{Saliba2009}. For a system with an FP laser diode, $\mathrm{Mode}_{\mathrm{int}}$ would have a much higher amplitude, resulting in a more complex mode structure. This leads to a higher instability of the ULNL, as described in Section \ref{4}.

The complete mode structure of an ULNL is shown in Figure \ref{fig:modesb}, which is calculated using Equation \ref{eq:2} with $r_{1}$, $r_{\mathrm{eff}}$ and the total optical length of all cavity parts. Compared to Figure \ref{fig:modesa} it becomes obvious that the most dominant part of the system is the ECDL: The grating with a reflectivity of $r_{\mathrm{ex1}}^2\approx0.15$ in combination with its frequency selective properties leads to single-mode laser operation (Figure \ref{fig:modesb} red dot). However, the additional delay line has a significant drawback: the small FSR makes the system susceptible to external influences. Small changes in the internal cavity length caused by fluctuations of electrical current or temperature, as well as variations in the external cavity length through fluctuations of air pressure or temperature shift the mode structure of each cavity part. This changes the combined transmittance profile, leading to variations in the amplitude of peaks near the main peak. If one of these peaks has higher gain than the actual emission peak, a mode jump occurs as the emission frequency jumps to the new maximum of $T$. Note that this approach to the modelling of mode behaviour of a laser resonator does not take into account any semiconductor physics effects.


\section{Experimental setup} \label{3}

\begin{figure}[b!]
\centering
    \subfigure{\includegraphics[height=0.6\textwidth]{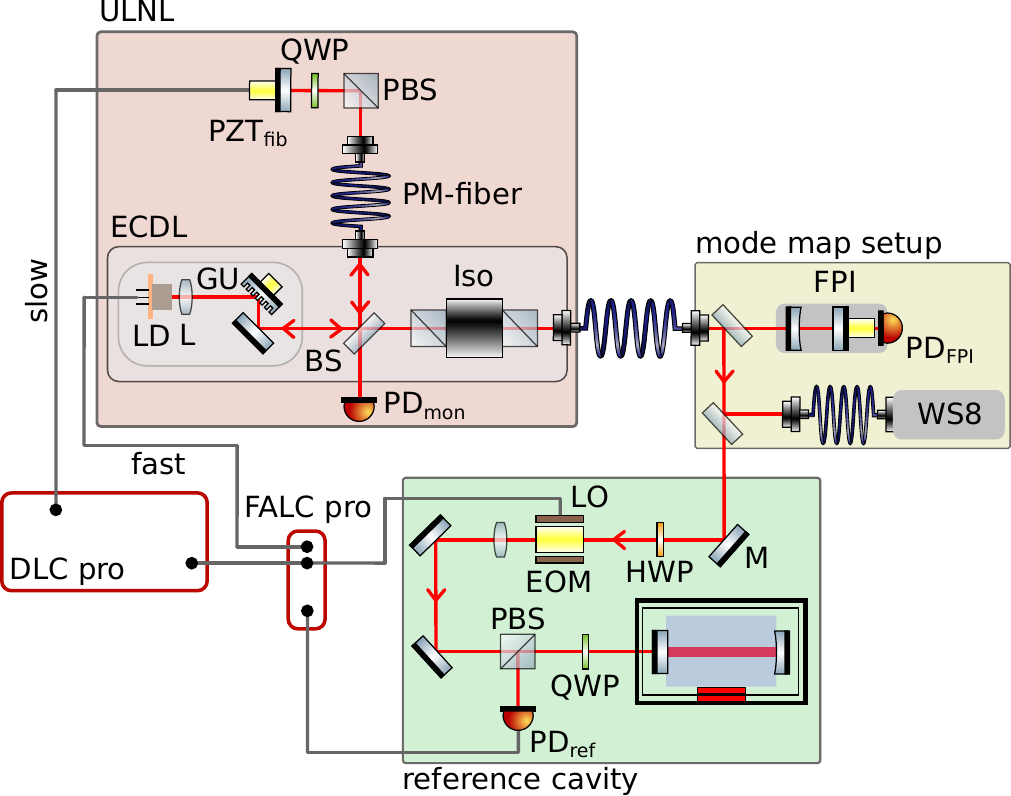}}
    \caption{\label{fig:setup}Schematic overview of the laser diode-based ULNL including electronics, the mode map measurement setup and the reference cavity setup. The component-library \cite{grafik} was used to create this graphic.}
\end{figure}

The experimental setup is shown in Figure \ref{fig:setup}. The light sources for our ULNL are indium gallium arsenide phosphide laser diodes. Two identical laser systems are set up for comparison, one with an FP laser diode and one with an AR laser diode. The ECDL is based on a TOPTICA DLC DL pro laser system in Littrow configuration \cite{Ricci1995}. The ECDL includes the grating (GU), which controls the cavity length $L_{\mathrm{ex1}}$ of the first external cavity. Behind the ECDL, a beam splitter (BS) with a splitting ratio of 90:10 (T:R) splits the light into two parts. The $10$\,\% port is coupled into the long external fiber cavity with an $1.5$\,m polarization-maintaining fiber (PM-fiber) \cite{Lin2012, Samutpraphoot2014}. The FSR of the fiber cavity is $\approx55$\,MHz, including the free space in front of the fiber. A silver mirror at the end of the fiber cavity creates, together with the GU, the second external cavity and reflects the light back to the laser diode. The mirror is placed on a piezoelectric actuator (PZT$_{\mathrm{fib}}$), which allows to change the length $L_{\mathrm{ex2}}$ of the additional cavity over multiple FSR. A combination of a quarter-wave plate (QWP) and a polarization beam splitter (PBS) is used to adjust the optical feedback power. A photodiode (PD$_{\mathrm{mon}}$) measures the level of reflected feedback power, which is crucial to calculate the feedback ratio $\beta$. Therefore, the reflection factors of the BS, the grating and the diode-coupling losses must be taken into account. The output of the $90$\,\% port of the BS passes through a $60$\,dB optical isolator (Iso) with a transmission efficiency of $90$\,\%. The light is coupled into an optical fiber with a coupling efficiency of $60$\,\%. The system generates $22$\,mW of optical power out of the fiber. The entire laser setup is placed on a transportable optical breadboard. In combination with a diode-based tapered amplifier system (TOPTICA DLC BoosTA pro), the fiber coupled output power can be increased up to $300$\,mW without an increase in FN (not shown in the figure).

For measuring the FN spectrum we detect the power spectral density $S_{\nu}(f)$ with a delayed self-heterodyne setup. Therefore, we separate the light into two parts: One part is frequency shifted using a $40$\,MHz acousto-optical modulator (AOM, AA Opto-Electronic MTS40-A3-750.850). The second part is delayed using a fiber with $20$\,m length, which leads to a delay of $\approx100$\,ns. Both beams are overlapped on a photodiode. An oscilloscope (Teledyne LeCroy HDO6104A) detects the beat signal and a software analyses the FFT to calculate the FN spectrum of the laser for Fourier frequencies between $300$\,Hz to $6$\,MHz \cite{Tsuchida2011, Schmidt-Eberle2023}.

The connection between two independent actuating parameters (e.g. the lengths of the two external cavities) and the behaviour of the longitudinal modes of a laser can be depicted in a two dimensional plot. These so-called mode maps are generated with the measurement setup shown in Figure \ref{fig:setup}. The light is split in two paths by a BS. A wavemeter (High Finesse, WS8) with a resolution of $200$\,kHz measures the frequency of the light. Simultaneously, a scanning Fabry-Pérot interferometer (FPI) with an FSR $=1$\,GHz and a $\mathcal{F} \approx 300$ is used to identify single-mode regimes of operation. Therefore the internal piezo of the FPI scans continuously the cavity length over several FSR, while the photodiode (PD$_{\mathrm{FPI}}$) behind the FPI records the transmitted intensity. When the laser is single-mode and resonant with the FPI, distinct sharp and narrow transmission peaks are detected with PD$_{\mathrm{FPI}}$. The distance between the peaks is constant. When the laser starts to run multi-mode, these main peaks first lose intensity, while additional small peaks become visible. These additional peaks originate from other excited longitudinal modes. If the additional peaks exceed a threshold of $> 2$\,\% compared to the main peak, the laser is recognized as multi-mode. This measurement is performed for any tuple of values of interest. We defined two independent variables, which are iterated through in two nested loops with specific step sizes, which are typically $\leq1$\,\% of the maximum allowed value. For each set of parameters a waiting period is implemented (settling time $\approx0.5$\,s) before the recording of the measured values. Each pair of values represent one point in the mode map. In single-mode regions the measured wavelength is encoded according to a color scale, while multi-mode regimes are depicted in white. The example of such a mode map is shown in Figure \ref{fig:modemap}; the details will be discussed in Section \ref{4}.

For laser frequency stabilization, the ULNL can be locked to the fundamental transverse mode of a plano-concave reference cavity, which has an FSR $=1.5$\,GHz and a $\mathcal{F} \approx 10000$. The Pound-Drever-Hall technique \cite{Pound1947, Drever1983} is utilized to establish the lock. To implement this technique, an electro-optic modulator (EOM, QUBIG PM7-NIR) in conjunction with a local oscillator (LO, TOPTICA Pound-Drever-Hall module) is utilized to generate $25$\,MHz sidebands. The control signal for the lock can be detected using a half-wave plate (HWP) and a PBS in reflection with a photodiode (PD$_{\mathrm{ref}}$). The TOPTICA locking electronics (DLC pro with lock option, FALC pro) are utilized to feed the control signal back to the laser system. Two actuators are employed to maintain resonance with the cavity. Fast fluctuations are corrected with the current of the laser diode, while the slow fluctuations are addressed using the piezo stack of the fiber cavity (PZT$_{\mathrm{fib}}$).


\section{System comparison and discussion} \label{4}

In order to ensure a fair comparison between the two ULNLs, both laser systems were characterized simultaneously. Also, all components of significant importance that impact laser performance, such as the grating GU, BS, and PM fiber, were thoroughly measured and selected beforehand. This ensures that the only difference between the two systems is the used laser diode type.

\subsection*{Linewidth reduction and influence of $\beta$}

\begin{figure}[t!]
\centering
    \subfigure[\label{fig:FNDa}]{\includegraphics[height=0.33\textwidth]{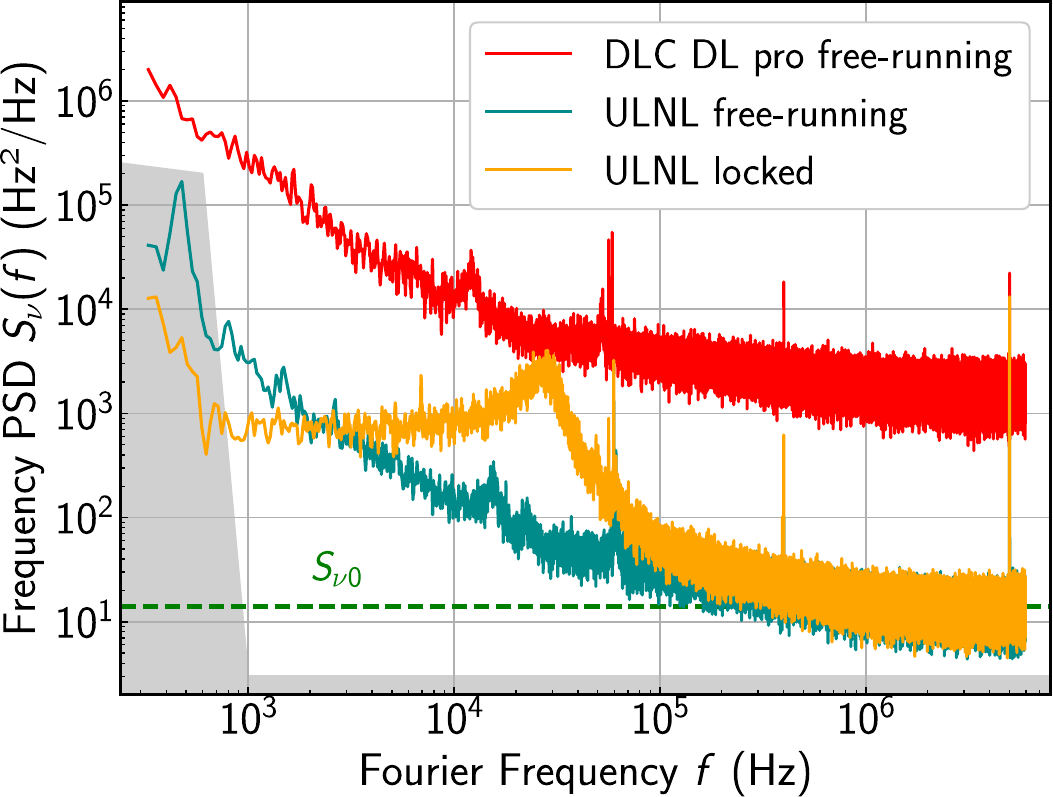}}\qquad
    \subfigure[\label{fig:FNDb}]{\includegraphics[height=0.33\textwidth]{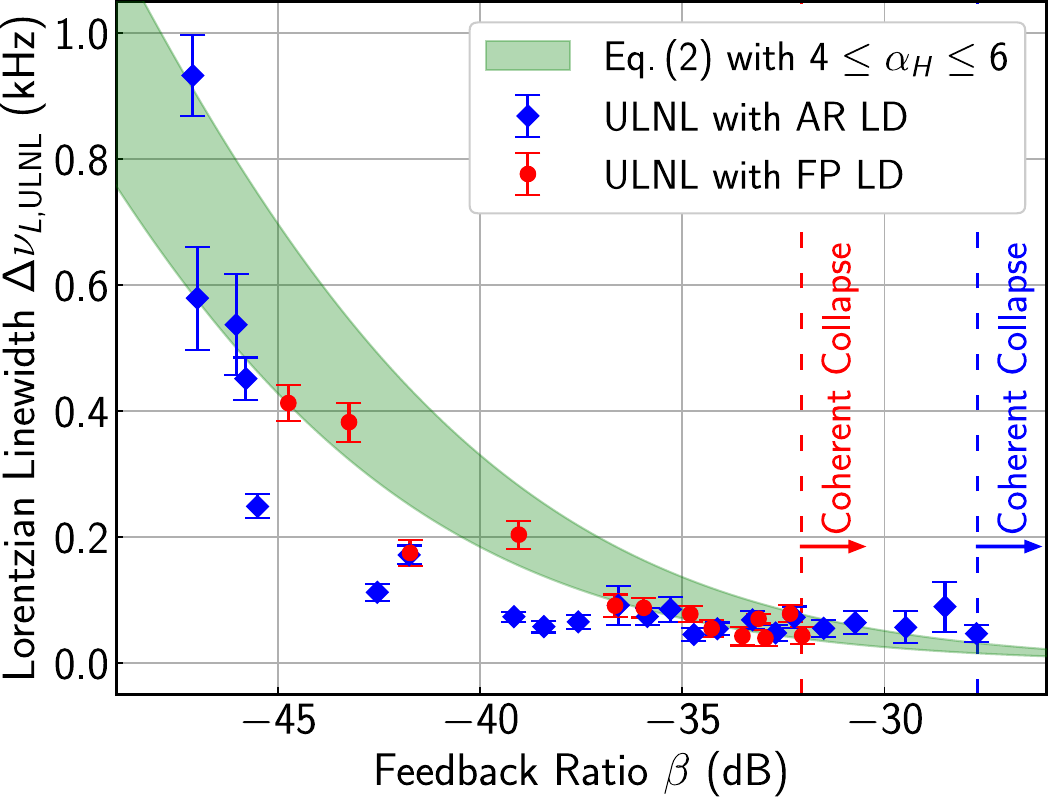}}\\
    \caption{(a) Comparison of FN power spectral density (PSD) traces for different kinds of ECDLs at $729$\,nm. "DLC DL pro free-running" (red) shows a typical FN spectrum for a DLC DL pro, "ULNL free-running" (cyan) is a DLC DL pro with additional optical feedback ratio $\beta=-35$\,dB from the fiber cavity, "ULNL locked" (orange) is the ULNL locked to the reference cavity setup shown in Figure \ref{fig:setup}. The grey area is the noise floor of the measurement setup. (b) Experimental linewidth $\Delta\nu_{L, \mathrm{ULNL}}$ for FP (red circles) and AR (blue diamonds) ECDLs as a function of fiber cavity feedback ratio $\beta$. The theoretical linewidth is calculated with Equation \ref{eq:6} (green) assuming $\alpha_{H}$ in a range of values: $4\leq\alpha_{H}\leq6$.}
\end{figure}

Figure \ref{fig:FNDa} presents FN measurements for different ECDLs. In terms of FN, the systems utilizing an AR laser diode and an FP laser diode exhibit similar behaviour. For clarity only FN power spectral density traces for the laser with an AR laser diode are shown. The "DLC DL pro free-running" (red) represents the typical FN of a commercially available free-running DLC DL pro laser at $729$\,nm \cite{Schmidt-Eberle2023}. Its characteristics demonstrate a transition from $1/f^n$ noise, $1 < n < 2$, around $100$\,kHz saturating to the level of white noise at $S_{\nu 0} = 2 \cdot 10^{3}$\,Hz$^{2}$/Hz. This results in a Lorentzian linewidth of $\Delta\nu_{L, \mathrm{ECDL}} \approx6$\,kHz according to Equation \ref{eq:5}. The "ULNL free-running" (cyan) depicts the FN of an ULNL with $\beta=-35$\,dB. The FN over the whole measurement range is around two orders of magnitude lower than the FN of the free-running commercial laser system. In the ULNL configuration, $S_{\nu 0}$ yields a Lorentzian linewidth of $\Delta\nu_{L, \mathrm{ULNL}} \approx0.1$\,kHz. When the system is frequency stabilized as described in Section \ref{3}, the resulting FN corresponds to the "ULNL locked" (orange) curve. The fact that the free-running ULNL is two orders of magnitude smaller in linewidth compared to the DLC DL pro allows for lock settings with weaker gain: The position of the servo bump can be tuned to a frequency of 30 kHz by adjusting the locking parameters in the FALC pro. This allows to keep the fast FN above $100$\,kHz at levels which are similar to the free-running ULNL and thus two orders of magnitude below a typical ECDL, which is an important property for the applications discussed in Section \ref{1}.

Figure \ref{fig:FNDb} illustrates the variation of $\Delta\nu_{L, \mathrm{ULNL}}$ as a function of $\beta$ for both ULNLs. Both systems exhibit a reduction in $\Delta\nu_{L, \mathrm{ULNL}}$ due to the presence of an additional fiber cavity. It is possible to achieve a linewidth of $\Delta\nu_{L, \mathrm{ULNL}} \leq 0.1$\,kHz with AR and FP laser diodes. However, the system employing the FP laser diode tends to require higher values of $\beta$ to achieve the same linewidth reduction as the system with the AR laser diode with $\beta<-37$\,dB. This can be attributed to the higher reflectivity of the partially reflecting front facet of the FP laser diode. It leads to increased reflection losses, resulting in reduced feedback entering the laser diode. Beyond a feedback ratio threshold of $\beta=-37$\,dB, the linewidth reduction for both systems becomes equal and reaches a minimum. For $\beta <-42$\,dB, the linewidth reduction decreases. Figure \ref{fig:FNDb} also depicts the feedback ratios at which both systems undergo a transition to multi-mode behaviour: for the AR laser diode it is at $\beta >-27$\,dB and for the FP laser diode it is at $\beta >-32$\,dB. This effect is known as coherence collapse \cite{Tkach1986}. In the literature the optical feedback regime for a single-mode operation with narrow linewidth is known as regime III. This regime is known to be independent of feedback phase and it provides stable narrow-line laser operation. This is exactly the regime in which we operate our ULNLs. It usually spans only a small range of feedback ratios. The pioneering work of Tkach and Chraplyvy \cite{Tkach1986} provided $-45 < \beta < -39$\,dB as boundaries. However, these values were measured only for one type of laser diode. Further works revealed that the boundaries of regime III can depend on several parameters \cite{Okoshi1980, Kane2005, Coldren2012}: the type of laser diode, larger output powers, longer cavity lengths and smaller laser coupling factors help to increase the width or shift the boundaries of regime III. In our case we see the boundaries are shifted for both laser diodes, which could be a result of the mentioned parameters. However, the transition to the coherent collapse happens at different feedback levels for AR and FP laser diodes. In summary the ULNL with an AR laser diode achieve $\Delta\nu_{L, \mathrm{ULNL}} = 0.1$\,kHz in a range of $\Delta\beta\approx 13$\,dB, while the ULNL with an FP laser diode achieve it within $\Delta\beta\approx 5$\,dB. The broader regime III makes the AR system more stable against feedback power fluctuations.

Equation \ref{eq:6} is used to theoretically describe the reduction of the linewidth (presented as a green shaded area in Figure \ref{fig:FNDb}). We calculate $\Delta\nu_{L, \mathrm{ULNL}}$ in a range of $4\leq\alpha_{H}\leq6$, because $\alpha_{H}$ changes its value depending on the exact position inside of a single-mode plateau \cite{Genty2002, Wenzel2021}. During the measurement the free-running laser drifts in frequency and changes the position inside the plateau, which leads to a change of $\alpha_{H}$. However, in general the trend of the theoretically predicted $\Delta\nu_{L, \mathrm{ULNL}}$ for an increasing $\beta$ fits to the measured data points reasonably well.

\subsection*{Mode stability of ULNLs}

\begin{figure}[b!]
\centering
    \subfigure{\includegraphics[width=0.9\textwidth]{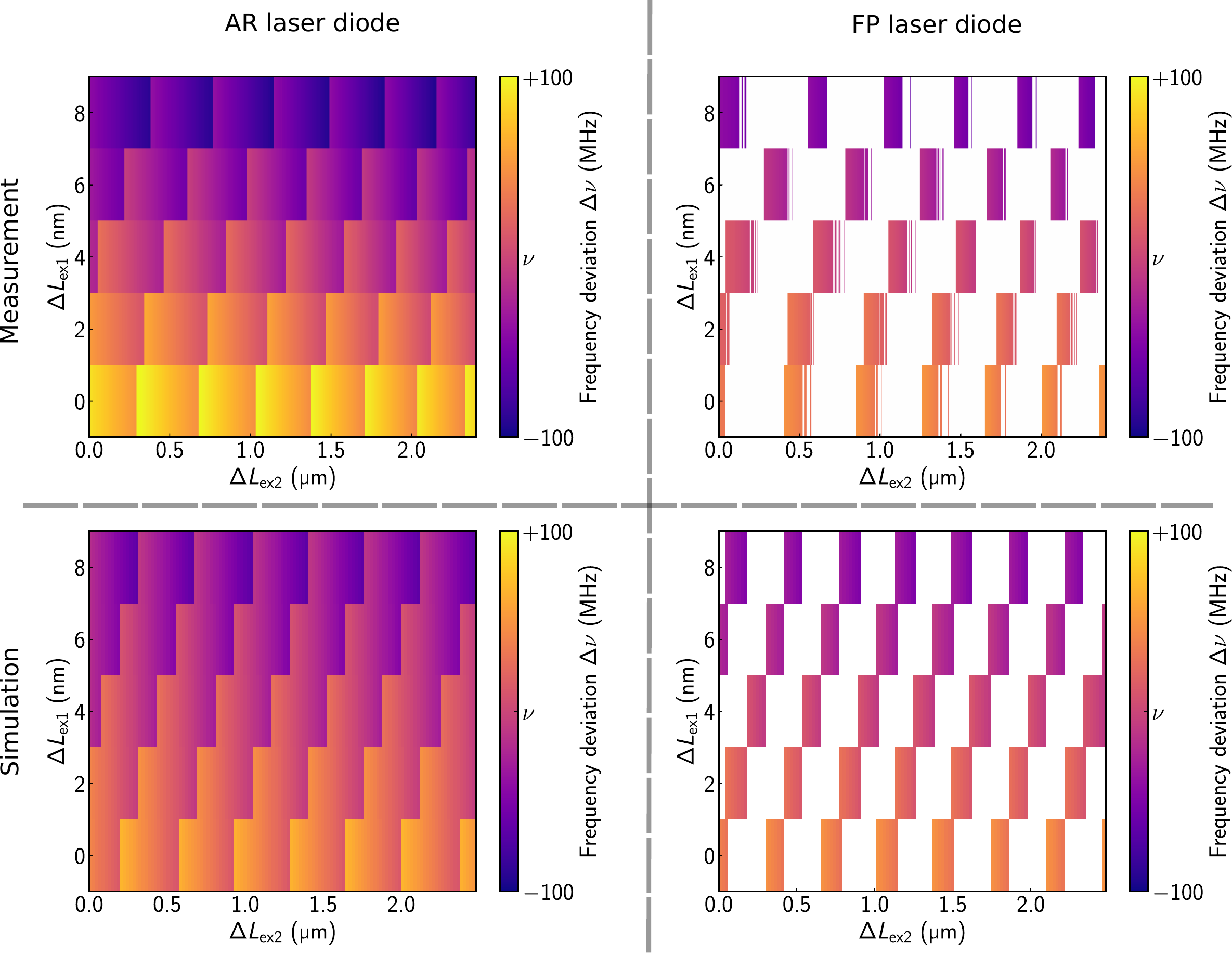}}
    \caption{\label{fig:modemap}Measured mode maps for ULNL with an AR laser diode and an FP laser diode compared to simulated mode maps for the two ULNLs based on Equation \ref{eq:2}. The two independent variables are $L_{\mathrm{ex1}}$ and $L_{\mathrm{ex2}}$. The white areas represent multi-mode regimes.}
\end{figure}

Significant differences between the two laser systems are observed in their mode behaviour. Mode maps are recorded following the procedure described in Section \ref{3} and are presented in Figure \ref{fig:modemap}.

During the measurement, changes in the semiconductor gain medium should be prevented from influencing the mode structure of the laser. Diode current or diode temperature, which change the diode properties, remain constant. For this reason, we concentrate on changes in the length of the passive components in the laser. One parameter in the measurement is PZT$_{\mathrm{GU}}$, which influences the length of first external cavity $L_{\mathrm{ex1}}$. The second parameter PZT$_{\mathrm{fib}}$ changes the length of the second external cavity $L_{\mathrm{ex2}}$. Both systems are set up with $\beta=-35$\,dB, because with this level of feedback both lasers achieve similar reduction of the linewidth (Figure \ref{fig:FNDb}). At the same time this value allows to stay sufficiently far away from the boundary to the coherent collapse for a given performance. We also used the theory discussed in Section \ref{2} to simulate the mode behaviour and to compare the model with the measurements. In order to distinguish between single-mode and multi-mode operation in the simulation, an empirical multi-mode detection function is integrated. For this purpose, the difference between the highest and second highest mode is determined. If this value falls below a threshold value, the operation is evaluated as multi-mode operation and the data point is then displayed in white.

The measurement starts with a calibration which brings both lasers into a single-mode regime of operation. In the left column of Figure \ref{fig:modemap}, the behaviour of the system employing the AR laser diode is depicted. Each mode map starts at the bottom left. First, the parameter PZT$_{\mathrm{GU}}$ is fixed, while the parameter PZT$_{\mathrm{fib}}$ changes in value, which leads to an increase of $L_{\mathrm{ex2}}$. The mode map demonstrates that single-mode operation is maintained throughout the whole measurement. The observed mode jumps correspond to the $\mathrm{FSR}=55$\,MHz of the ULNL. Secondly, the PZT$_{\mathrm{GU}}$ makes a step to the next value, the length of the first external cavity changes by around $\Delta L_{\mathrm{ex1}} = 2$\,nm. Correspondingly, the emission frequency of the ECDL changes by a few MHz. For this value, the same change of $\Delta L_{\mathrm{ex2}}$ leads to the same mode hop behaviour as before. In total it is possible to set every desired $\nu$ in a range of $\Delta\nu \pm 100$\,MHz in the given parameter range. The same mode behaviour is observed in the simulation. Mode hops arising between longitudinal jumps of the second cavity matching its FSR are observed. The range of simulated $\Delta \nu$ over the whole parameter range is a few MHz lower than for the measured data, which we attribute to a drift of the center frequency during the measurement. Nevertheless the simulations describe the mode behaviour of the AR laser diode-based ULNL sufficiently well to predict the influence of changing cavity lengths.

In contrast, the mode maps of the laser system employing the FP laser diode reveal significantly worse single-mode behaviour (Figure \ref{fig:modemap} right column). The laser regularly switches between single-mode and multi-mode operation while scanning the length of the second external cavity $L_{\mathrm{ex2}}$. The multi-mode regions are approximately twice as large as the single-mode regions. The single-mode regions are reduced to $20$\,MHz tuning range and are thus significantly smaller compared to the system with the AR laser diode. This behaviour repeats for every setting of the length of the first external cavity $L_{\mathrm{ex1}}$. This represents a crucial difference compared to the AR laser diode. However, single-mode areas with limited tuning can also be found. The simulations show qualitative agreement with the measurements, but differ in the size of the multi-mode areas. This can be explained by the simple structure of the simulations, which only take into account the effect of changes in the cavity lengths, while the physics of the semiconductors dynamics is only considered empirically. The operation of a real laser is more complex than our simulation, but our simple model is well suited to qualitatively describe the mode structure of complex ECDLs.

Our investigation reveals that in the case of the AR laser diode, regime III is independent of the feedback phase. Tuning the laser cavities affects the laser wavelength as expected but does not lead to multi-mode operation. We believe that this behaviour originates from the simpler cavity geometry of the AR laser diode-based ULNL, effectively lacking one cavity compared to the FP laser diode system. The original feedback regime theory was developed for a three mirror case corresponding more closely to the AR system.


\section{Conclusion} \label{5}

In this study, we presented an investigation of the behaviour of two ULNLs based on different types of laser diodes. Specifically, we modify TOPTICA DLC DL pro ECDLs by integrating an additional fiber cavity. This approach combines the advantageous optical FN characteristics of TiSa laser systems with the advantages of ECDLs, such as flexibility in the wavelength regime, compact size and cost effectiveness. Our focus lies on $729$\,nm, which is relevant for a specific calcium ion transition utilized in quantum computing, quantum simulation and optical clock applications that demand stringent FN requirements in the MHz regime.

Our study encompasses two distinct types of measurements. Firstly, we analyze the FN performance utilizing a self-heterodyne measurement setup. Secondly, we examine the behaviour of the ULNL single-mode operation by measuring the mode maps of the system as a function of the external cavity lengths. In terms of FN measurement, the results obtained using both laser diode types demonstrate almost identical performance. This outcome is attributed to the similarity in the total length of the cavity employed in both systems. Differences between the two systems concerning the onset of coherence collapse and the feedback ratio required to achieve the same linewidth suppression can be attributed to the additional reflective surface in the FP system.

However, the single-mode behaviour exhibits significant differences between the two laser diode types: The ULNL based on an AR laser diode exhibits distinct regions for single-mode operation. In contrast, the system employing an FP laser diode frequently transitions between single-mode and multi-mode states, resulting in a less predictable behaviour. These findings underscore the critical importance of careful consideration when selecting the type of laser diode for achieving stable and reliable operation of ULNL systems. However, laser diodes with an AR coating are often not available. In this case, the ULNL with an FP laser diode can also be used. However, more careful tuning and control is required to keep the laser with an FP laser diode in a single-mode state.

The implications of our results are particularly significant for quantum applications, as they facilitate while maintaining a compact and robust system. This holds particular relevance for modern highly integrated experiments, where the combination of precision, compactness and reliability is of utmost importance.


\section*{Funding}
We gratefully acknowledge funding by the European Union’s Horizon 2020 research and innovation program AQTION (grant agreement No. 820495), the Federal Ministry of Education and Research (BMBF) project QRydDemo (grant agreement No. 13N15635), the German Research Foundation (DFG) projects DQ-mat (ID. 274200144 – SFB 1227) and terraQ (ID. 434617780 – SFB 1464) and Germany’s Excellence Strategy QuantumFrontiers (ID. 390837967 - EXC-2123).

\section*{Acknowledgements}
We thank Hans Wenzel for helpful discussions and Heather Partner for careful review of the manuscript.

\pagebreak
\bibliographystyle{unsrt}
\bibliography{main}

\end{document}